\documentstyle[aps,prb,twocolumn,graphicx]{revtex}

\newcommand{\cm}{cm$^{-1}$}
\newcommand{\cgo}{CuGeO$_3$}
\newcommand{\nvo}{NaV$_2$O$_5$}
\newcommand{\telep}{Sr$_{14-x}$Ca$_x$Cu$_{24}$O$_{41}$}
\newcommand{\titel}{LOW DIMENSIONAL CORRELATED SYSTEMS}

\begin{document}
\draft
\title{\titel: \cgo\ AND \nvo}
\author{P.~H.~M. VAN LOOSDRECHT}
\address{II. Physikalisches Institut der RWTH-Aachen, Templergraben 55,
            52056 Aachen, Germany}
\author{J. ZEMAN, G. MARTINEZ}
\address{Grenoble High Magnetic Field Laboratory MPI-FKF/CNRS,
         25 Avenue des Martyrs, F-38042 Grenoble Cedex 9, France}
\author{M.~J. KONSTANTINOVI\'C}
\address{MPI-FKF, Heisenbergstr.~1, D-70569 Stuttgart, Germany}
\author{A. REVCOLEVSCHI}
\address{Laboratoire de Physico-Chimie des Solides, Universit\'{e} de Paris-Sud,
            B\^{a}timent 414, F-91405 Orsay, France;}
\author{Y. UEDA}
\address{Institute for Solid State Physics, The University of Tokyo, 7-22-1
            Roppongi, Minato-ku, Tokyo 106, Japan.}

\date{Submitted to: Proceedings of the international conference on Low Energy Electrodynamics in Solids (LEES99),
      June 21-25 1999, P\'ecs, Hungary.}
\maketitle

\begin{abstract} Some of the properties of the
low-dimensional electronically correlated materials \cgo\ and
\nvo\ are discussed. The emphasis lies on recent results obtained
using Raman scattering and optical absorption spectroscopy as a
function of temperature, magnetic field and hydrostatic pressure.

\end{abstract}

\section{INTRODUCTION}

Low-dimensional correlated systems show a wide variety of
physically interesting and unusual properties. Well known in this
field are the copper-oxides showing high temperature
superconductivity \cite{BED86}, and the low-dimensional spin
systems such as the spin-Peierls compound \cgo\ \cite{HAS93}, the
Sr-Cu-O ladder compounds \cite{DAG96}, the superconducting
chain/ladder compound \telep\ \cite{UEH96}, and the recently
discovered charge-ordered compound \nvo\ \cite{ISO96}. One of the
interesting aspects of one-dimensional spin-chains with isotropic
interactions is that the presence of a continuous symmetry
prevents spontaneous ordering \cite{MER66}. In these systems
ordering may occur only through coupling to other degrees of
freedom, such as lattice or charge excitations. The transition to
a singlet ground state in \cgo\ is, for instance, a direct result
of the spin-phonon coupling in this system.

In this contribution, a few of the fascinating properties of
low-dimensional correlated systems --in particular of \cgo\ and
\nvo-- are briefly discussed. Section II introduces the
compounds \cgo\ and \nvo. Section III discusses the
nature of spin-excitations and the occurrence of bound states in
low-dimensional spin systems. Section IV addresses the
($H-T$) phase diagram of, in particular, \cgo. Finally, section
V discusses charge-ordering in \nvo, and presents some
ideas on low-energy charge excitations in this compound.

\section{\cgo\ AND \nvo \label{cgonvo}}

\subsection{The spin-Peierls compound \cgo}

\cgo\ is the first example of an inorganic compound \cite{HAS93}
showing a spin-Peierls transition \cite{BRA83}. Characteristic
fingerprints of a spin-Peierls transition are the formation of a
singlet ground state, evidenced by a vanishing magnetic
susceptibility; the formation of a superstructure, as evidenced by
the appearance of superlattice reflections in diffraction
experiments \cite{POU94,HIR94}; and the opening of a spin gap in
the magnetic excitation spectrum, as evidenced in, for instance,
inelastic neutron scattering experiments \cite{NIS94}.

The spin-chains in \cgo\ are formed by unpaired $d$-electrons on
Cu$^{2+}$\ ions which are magnetically linked into chains along
the orthorhombic $c$-direction by an almost 90$^{\rm o}$\ Cu-O-Cu
super-exchange path. The chains are separated from each other by
GeO$_4$\ units. The existence of weak inter-chain couplings makes
this compound quasi-one-dimensional ($J_c\approx120$~K,
$J_b\approx16$~K, $J_a\approx2$~K) \cite{NIS94}. In addition,
there is a substantial next-nearest-neighbor interaction
($J_{nn}\approx 160$~K, $J_{nnn}\approx40$~K) \cite{RIE95,CAS95}.
The transition temperature in \cgo\ is $T_c\approx14$~K, and the
gap in the magnetic excitation spectrum is found to be $2.1$~meV
\cite{NIS94,REG96}.

\subsection{The charge-ordered compound \nvo}

Initially, \nvo\ was thought to be a spin-Peierls compound with
$T_c\approx34$~K \cite{ISO96}. Also in this case, an exponentially
vanishing susceptibility was observed, as well as the formation of
a superstructure and the opening of a spin-gap
($\Delta\approx10$~meV \cite{YOS98}). Moreover, the
temperature-dependent susceptibility in the high temperature phase
exibits the typical Bonner-Fisher behaviour \cite{BON64} expected
for an antiferromagnetic $S=1/2$\ spin-chain \cite{ISO96}. Recent
experimental findings \cite{OHA97,SCH99,GRO99} suggest, however,
that this compound is not a spin-Peierls compound but rather a
charge-ordered compound [18-22].

At high temperatures, \nvo\ may be considered as a quarter-filled
ladder system \cite{SMO98}. The rungs ($\parallel a$) of the
ladders ($\parallel b$) are formed by two Vanadium ions (VO$_5$\
pyramids) with intermediate valence 4.5+. The V ions are connected
through 180$^{\rm o}$\ V-O-V bonds. The $a-b$\ ladder layers are
separated, in the $c$-direction, by Na ions. At high temperatures
spin-chains are formed by the rungs, each containing one unpaired
electron. At low temperature a charge-ordering occurs, where the
two V ions on a single rung become inequivalent, {\it i.e.} they
now have different ($4.5\pm\delta$) valence states. The ordering
occurs most likely in a zig-zag pattern along the ladders.

\section{SPIN EXCITATIONS AND BOUND STATES\label{spinex}}

\begin{figure}[htb]
\begin{center}
\includegraphics[width=80mm,clip=true]{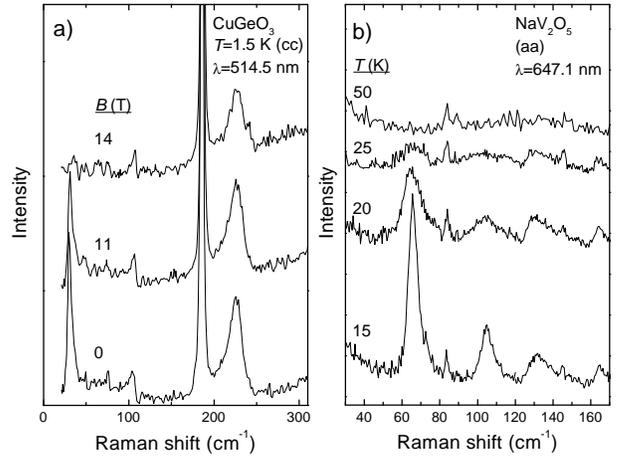}
\end{center}
\caption{\label{fig1}
a) Polarized ($cc$) Raman spectra of \cgo\ at $T=1.5$~K for $H=$0, 11,
   and 14 T \cite{pvl97a}.
b) Polarized ($aa$) Raman spectra of \nvo\ for $T=$\ 15, 20, 25 and
   50 K \cite{KON99}}
\end{figure}

Spin excitations in antiferromagnetic compounds are classically
described by spin-wave theory. In quantum spin systems, however,
this simple theory is no longer sufficient. Instead of having a
dispersion branch describing well-defined spin waves, one now has
spinons as elementary excitations. These $S=\frac{1}{2}$\ spinons
do not occur as single particles, but rather form a two-particle
triplet excitation continuum \cite{mul81}. In addition to that,
the interaction between these triplets may lead to the formation
of bound states and additional continua which may be of singlet,
triplet or even quintuplet nature. Such bound states have indeed
been observed. In \cgo\ it reveals itself \cite{KUR94,PVL96a} as a
well defined mode at 30 \cm\ (see figure \ref{fig1}a)). Its
singlet nature is evidenced by the lack of field dependence in the
dimerized phase ($H<12.5$~T). In the high field phase
($H>12.5$~T), the bound state mode either moves to lower energy or
disappears all together.

In \nvo, the situation is a bit more complex. Also here, one finds
newly activated modes in the low-temperature phase
\cite{LEM98,KUR98,KON99} (see figure \ref{fig1}b). In this case,
however, several modes are observed -- at 66, 106, and 132 \cm\ --
that have been interpreted as singlet bound states \cite{LEM98},
as low-lying Vanadium $d$-level excitations \cite{KON99}, or as
folded phonon excitations with a renormalized energy and intensity
due to spin-phonon interactions \cite{PVL99d}. Presently, the
exact origin of these low-energy modes is still under intense
debate. The availability of inelastic neutron scattering data, as
well as an improved understanding of, in particular, the low
energy dynamics in \nvo\ appear to be a prerequisite for a full
understanding of these low-energy modes.

\section{THE ($H-T$) PHASE DIAGRAM \label{phased}}

\begin{figure}[htb]
\begin{center}
\includegraphics[width=7cm]{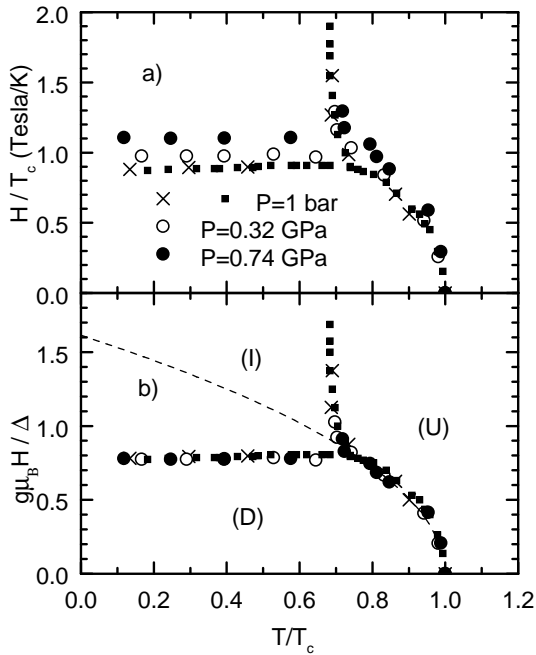}
\end{center}
\caption{\label{fig2}
Phase diagram of CuGeO$_3$\ for different
pressures plotted versus $T/T_{c}$ \cite{PVL99}; a) scaling of
$H$\ with $T_{c}$\ and b) scaling of $H$\ with $\Delta_{ST}$. The
solid squares are data obtained at 1~bar by
magnetostriction/thermal expansion measurements
\cite{PVL98b}.}
\end{figure}

One of the characteristic properties of a spin-Peierls compound is
its ($H-T$) phase diagram (see figure \ref{fig2}) consisting of a
uniform phase ($T>T_{c}$), a dimerized phase ($T<T_{c}$; $H<H_c$),
and a modulated phase for high fields ($T<T_{c}$; $H>H_c$). For
\cgo, this general structure is experimentally well supported (see
for instance \cite{PVL98b}). The full phase diagram of \nvo\ has
not yet been determined. It has been shown, however, that the
phase transition is rather insensitive to applied fields
\cite{SCH99}, strongly supporting the idea that this is not a
spin-Peierls, but rather a charge-ordered compound.

Standard theory predicts the phase diagram of a spin-Peierls
compound to scale as $H/T_c$\ and $T/T_c$\ \cite{CRO79a}. Optical
absorption experiments on \cgo, in which $T_c$\ was varied by
applying hydrostatic pressure, have shown that this compound does
not follow this scaling (figure \ref{fig2}a). Instead a
$H/\Delta$, $T/T_c$\ type of scaling has been found (figure
\ref{fig2}b), where $\Delta$\ is the spin-gap (about 2.1 meV in
\cgo\ \cite{NIS94}) at zero magnetic field. This observed
$H/\Delta$\ scaling may not be so surprising: the gap state has a
triplet character, and therefore shows a linear splitting in a
magnetic field. A phase transition is expected when the lowest
energy triplet state energetically crosses the ground state. The
surprising observation is, however, that the first-order phase
transition into the incommensurate phase occurs already when the
gap from the ground state to the lowest triplet state is still
about  $\Delta/4$, {\it i.e.} at much lower magnetic fields than
intuitively expected. These observations call for a theory going
beyond the standard Cross-Fisher theory. They also call for a
reexamination of the organic spin-Peierls compounds, where, in
particular, an accurate determination of the spin-gap value is of
interest. An initial test for $H/\Delta$\ scaling, based on
spin-gaps determined from susceptibility experiments, has indeed
already shown that these compounds may show a similar scaling
behaviour \cite{PVL99}.

\section{CHARGE ORDERING AND EXCITATIONS IN NaV$_2$O$_5$ \label{co}}

The charge-ordering transition in \nvo\ occurs at $T=34$~K. This
relatively low transition temperature implies that there should be
a low energy scale for the charge dynamics of \nvo. At first
sight, the interactions determining the charge dynamics are the
intersite Coulomb interactions, and the intersite hopping terms.
These interactions, however, have typical energies of about 0.5
eV. It is not easy to see how these high energies may lead to a
low-energy scale.

\begin{figure}[htb]
\begin{center}
\includegraphics[width=7cm]{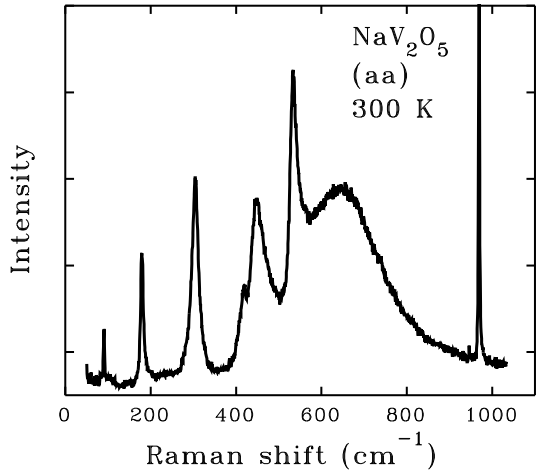}
\end{center}
\caption{\label{fig3}
Room temperature $(aa)$-Polarized Raman
spectrum of \nvo\ \cite{PVL99bb}.}\vspace{5mm}
\end{figure}

Experimentally, there is some evidence from optical spectra that
such a low energy scale indeed exists \cite{KON99,DAM98,FIS99}.
Raman scattering, for instance, shows a broad band in the $(aa)$\
spectra, already at room temperature (see fig. \ref{fig3}). One of
the origins of this scattering continuum might be high energy spin
excitations activated through an exchange coupling process,
similar to that observed in \cgo\ \cite{PVL96a}. The energy of the
observed band is, however, incompatible with this. From the
exchange interaction $J=560$~K \cite{ISO96} of \nvo, one would
expect two-magnon scattering to occur around 1000 \cm\ and not
around 500 \cm, as observed. Furthermore, one would expect this
scattering to occur in a $(bb)$-polarized configuration ({\it
i.e.} with light polarizations along the dominant exchange
direction), whereas the low energy continuum is observed in
$(aa)$-polarization only ({\it i.e.} polarizations along the
rungs).

Presently, the low energy scale is still an open question. There
is a simple single ladder model which gives some first insight
into this problem \cite{PVL99bb}. In this model, the only
parameters are the Coulomb interaction, $V$, between vanadium ions
on different rungs along the ladder, and the hopping, $t$, between
ions on a single rung. The Hamiltonian for this system is given by
\begin{equation}
H=t\sum_i{(c^{+}_{i,l}c_{i,r}+c^{+}_{i,r}c_{i,l})}+V\sum_{i,s=l,r}
{\rho_{i,s}\rho_{i+1,s}},\label{ham}
\end{equation}
where $c_{i,s}$ ($c^{+}_{i,s}$) are electron annihilation
(creation) operators for site $s$\ on the $i^{th}$\ rung, and
$\rho_{i,s}$\ are the charge densities on these sites. As long as
$|V|<t$, this Hamiltonian leads to a homogeneous charge
distribution on the rungs, {\it i.e.} the average position of the
electron on the rung is at the center. Starting from this
equilibrium state, it is clear that the {\it classical} elementary
excitations in such a system are charge waves involving small
displacements of the average electron position. The dispersion of
these excitations is easily derived from eq. (\ref{ham}) to give
$E_k=2t\sqrt{(1+(V/t)\cos(k\cdot x))}$, where $k$\ is the momentum
along the ladder, and $x$\ is in the ladder direction. At $k=0$\
these are relatively high energy excitations with uniform charge
displacements along the rungs (possibly observable in $a$
polarized infrared spectra \cite{DAM98}). The excitations at
$k=\pi/a$\ have an alternating charge displacement, and have a low
energy provided that $V\sim t$. These may indeed be Raman active,
in a $(aa)$-configuration through a two-particle process involving
states with momentum $\pm k$.

For $V>t$, the Hamiltonian eq. (\ref{ham}) leads to a zig-zag
charge-ordered phase with a charge disproportionately
$\delta=\sqrt{1-(t/V)^2}$. The typical low-energy excitations
would then be charge solitons -- which can be envisioned as kinks
in the zig-zag pattern -- with typical energy $(V-t)^{3/2}$
\cite{SHE99,RAJ82}. Provided that $V$\ is of the order of $t$,
this indeed gives again a low-energy scale.

The fact that \nvo\ charge-orders, implies that $V\sim t$\ for the
present simple model eq. (\ref{ham}). So, within the above model
one can easily understand the occurrence of low-energy charge
waves and charge solitons. The question is, however, how the
charge order is stabilized. Any small variation of either $V$\ or
$t$\ would directly lead to drastic changes in the charge
dynamics, as well as to the phase transition itself. Another
problem is that eq. (\ref{ham}) considers a single chain only.
This approach is not ideal of course. The presence of long-range
Coulomb interactions and inter-ladder hopping terms requires
inclusion of neighboring
ladders too [18-22] 
Hopefully more elaborate models can resolve these problems, and
give a full understanding of the charge-ordering and the-low
energy charge dynamics in \nvo.

\section{CONCLUSION}

Low-dimensional, correlated spin and charge systems show a wide
variety of intriguing physical phenomena. Many of these phenomena
may be understood using one-dimensional models, which theory can
handle quite well. Many of the interesting phenomena arise due to
the interactions between the various degrees of freedom such as
the spin-Peierls transition in \cgo, and the charge-order
transition and formation of a singlet ground state in
\nvo.\vspace{5mm}

\acknowledgements The GHMFL is ``Laboratoire
conventionn\'{e} \`{a} l'UJF et l'INPG de Grenoble''. The
Laboratoire de Physico-Chimie des Solides is a "Unit\'{e} mixte de
Recherche n$^{\circ}$ 8648". J. Z. acknowledges partial support
from the EC through grant ERBCHGECT 930034. M.K. thanks the Roman
Herzog-AvH foundation for support. This work is partially
supported by the DFG through SFB341.

\end{document}